\documentclass[12pt]{iopart}
\usepackage{iopams}  
\usepackage{graphicx}
\usepackage{bm}
\usepackage{color}
\begin{document}

\title[Proposal for test of the gravity of electrons and positrons]{Proposal for the realization of Santilli's comparative test on the gravity of electrons and positrons via a horizontal supercooled vacuum tube}

\author{Victor-Otto de Haan}
\address{BonPhysics B.V., Laan van Heemstede 38, 3297 AJ Puttershoek, The Netherlands}
\ead{victor@bonphysics.nl}
~\\ \vspace{0mm} \hspace{24mm}{\it date}: \today

\begin{abstract}
A proposal for the realization of Santilli's comparative test of the gravity of electrons and positrons via a horizontal supercooled vacuum tube is described. Principle and requirements are described concerning the sources, vacuum chamber electromagnetic shielding and pressure and position sensitive detector. It is concluded that with current technology the experiment is perfectly feasible. 
\end{abstract}

\pacs{14.60.Cd;29.30.-h;07.05.Fb 
~ \\ ~ \\ {\it Keywords}: Anti matter gravity; Equivalence principle; Experimental proposal
~ \\ ~ \\ {\it Under partial support by The R. M. Santilli Foundation, Grant Number RMS-AM-4673rs82810}
~ \\ ~ \\ {\it Presented at ICLATIP-3 Kathmandu University Nepal, January 6, 2011}
}
\maketitle
\newpage 
\section{Introduction}
Although the equivalence principle is well established for neutral bulk matter~\cite{Eotvos},\cite{Roll} and neutrons~\cite{Dabbs},\cite{Koester},\cite{Littrel} it has no experimental verification for charged elementary particles or antimatter. 

Even the gravitational mass of the electron has not been measured. Although there has been an attempt to measure the gravitational mass of electrons in the 1960's by Witteborn and Fairbank~\cite{Witteborn},\cite{Witteborn2}, this experiment was inconclusive. The goal of this experiment was to determine the gravitational force on both electrons and positrons, but is was only performed with electrons yielding a result disputed in literature. The experiment was not repeated with positrons due to lack of an adequate positron source~\cite{Darling}. The primary cause of the failure of the experiment is the magnitude of the effect, comparable to the force on a elementary charge due to an electric field of $5.6\times 10^{-11}$ V/m, corresponding in magnitude to the force repelling two unshielded electrons 5~m apart in vacuum. All electric fields must be controlled within at least an order of magnitude better accuracy.

Efforts are underway to measure the equivalence principle for neutral antimatter at CERN~\cite{Drobychev},\cite{Doser},\cite{Kellerbauer} and Fermilab~\cite{Cronin},\cite{Kaplan} to avoid the problems associated with the charge of the particle. However, it is argued that the equivalence principle for matter or antimatter could be different from the one for charged elementary particles~\cite{Santilli01},\cite{Goldman1987} so that an experiment with electrons and positrons is still called for.

Since the first attempts of Witteborn~\cite{Witteborn} to measure the gravitational mass of an electron much effort has been invested in the study of the experimental difficulties reducing the electric field to theoretical acceptable limits. First, the focused changed from positrons to anti protons~\cite{Goldman1982} due to the large inertial mass difference between the elementary particles. Later after a 1996 workshop on antimatter gravity and anti hydrogen spectroscopy~\cite{Workshop1997} the focused changed again to neutral antimatter. The reason for this was the problem posed by the so-called {\it patch-effects}~\cite{Darling}. These effects were assumed to render the measurements with positrons and even with anti protons impossible. 

However, Witteborn and Lockhart have always maintained that the patch-effects were somehow shielded after cooling to a temperature of 4.2~K~\cite{Witteborn2},\cite{Lockhart},\cite{Lockhart2}. A possible shielding mechanism of the patch-effect was observed by Rossi~\cite{Rossi} and a patch-effect reducing with temperature and surface treatment has been observed over a metal surface~\cite{Labaziewicz}. Also Dittus~\cite{Dittus}, proposing a gravity experiment in space, argues that with modern techniques the patch-effect can be reduced significantly.  
  
The above shows the need for a comparison of the gravitation on electrons and positrons and addressed why until now this has not been performed. In view of the recent technological developments of surface treatment these limitations can now be overcome and the experiment in a free horizontal flight in a high vacuum tube as first proposed by Santilli~\cite{Santilli01} and its principles worked out by Mills~\cite{Mills} can now be performed with small technological risks.

In the following first the principle of the experiment is lined out, then the the several components are highlighted and finally the conclusions are given.

\section{Principle}
\begin{figure}[t]
\begin{picture}(300,160)
\put(100,0){\scalebox{0.9}{\includegraphics{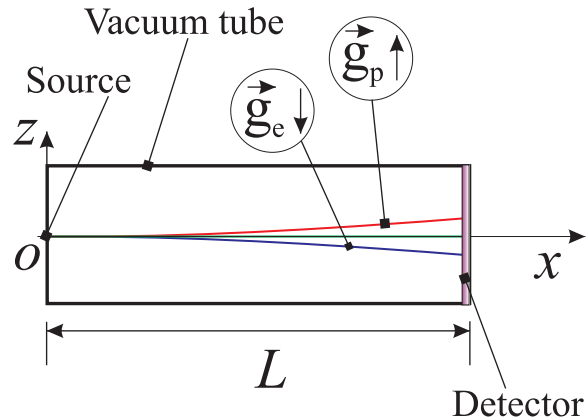}}}
\end{picture}
\caption{\label{fig1} Principle set-up of Santilli's comparative test of the gravity of electrons and positrons.}
\end{figure}

The principle of Santilli's comparative test of the gravity of electrons and positrons is shown in figure~\ref{fig1}. At one end of a well-shielded horizontal vacuum tube an electron or positron is released with a horizontal velocity, $v$. The particle moves through the vacuum tube until it reaches the other end at a distance $L$ and it is detected with a position sensitive detector. During the flight the particle experiences a constant gravitational acceleration, $\vec{g}_e$ or $\vec{g}_p$. The deflection at the end of the flight path is simply given by 
\begin{equation}
\Delta z_{e,p} = g_{e,p} \frac{ t^2}{2}
\end{equation}
where $t$ is the time the particle needs to reach the detector after is has been released at the source. This is called the time-of-flight. 

The deflection of the particle is proportional to the gravitational force so that measuring the deflection is sufficient to determine its sign. For neutral matter this set-up can be easily realized and with some more effort the same principle has been used to detect the gravity effects on neutrons~\cite{Dabbs},\cite{Koester}. 

However, the measured deflection also depends on the time-of-flight, which is simply given by $L/v$. {\it Hence, the deflection is inversely proportional to the (horizontal) kinetic energy of the particle.} The particle source will typical emit particles with some velocity distribution, hence the deflection is smeared out. This can be prevented by measuring the time-of-flight using a pulsed source. In that case the deflection of the particles is proportional to the square of the time-of-flight. 

\begin{figure}[b]
\begin{picture}(300,160)
\put(60,0){\scalebox{0.8}{\includegraphics{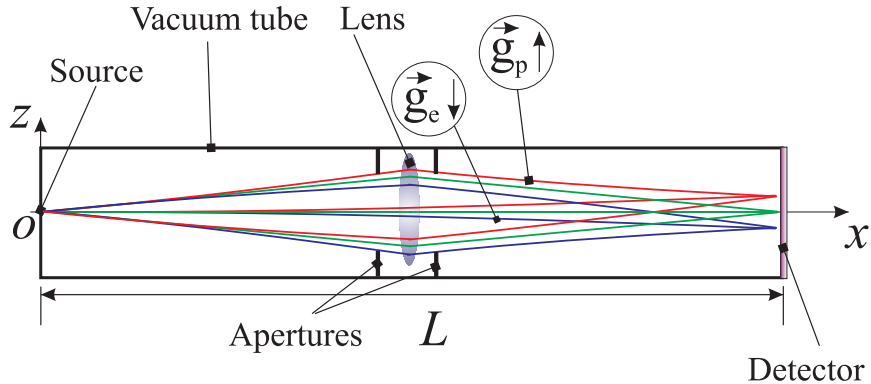}}}
\end{picture}
\caption{\label{fig2} Principle set-up of Mills's adaptation of Santilli's comparative test of the gravity of electrons and positrons.}
\end{figure}

Another assumption in the above reasoning was that the particles were emitted horizontally. With a typical particle source this direction will have some final spread around the horizontal, which again results in smearing out of the deflection. For neutral matter this is overcome by applying a diaphragm system to direct and collimate the particle beam. As Mills~\cite{Mills} has shown for charged particles a diaphragm system can be replaced by a focusing system and a suitable aperture system in the middle of the flight path. This relaxes the requirements for particle source strength quite a bit as a much larger divergence can be tolerated. With the focusing lens the source is imaged on the detector reducing the smearing out of the deflection. This is schematically shown in figure~\ref{fig2}. For a lens to work appropriate (with as small as possible aberrations) the lateral dimensions should be some two orders of magnitude smaller than the longitudinal dimensions (par-axial approximation). 

Another experimental feature that Mills incorporates is to reverse the flight direction keeping all other experimental conditions unchanged. The average of the four deflections is much less sensitive to remaining electric and magnetic stray fields and equal to 
\begin{equation}
\left\langle \Delta z \right\rangle= (g_e + g_p) L^2/v^2
\end{equation}
Hence both sides of the vacuum tube must provide sources of electrons and positrons and detectors of the same. This also limits the possibilities of the focusing system to a symmetrical one, with a magnification of 1. In the following sections some details on the main components are given.  

\section{Components}
\subsection{Electron and positron sources}

The main requirements for the electron and positron sources needed for this experiment can be inferred from figure~\ref{fig3}. To have a good compromise between maximal kinetic energy and minimal flight-path, the available source area must have a height of some 100 $\mu$m and a length of the order of a centimeter. The length can not be larger because then the focusing properties of the lensing system will be imparted. The height can not be larger as then too small kinetic energies would be needed. The kinetic energies needed are of the order of 1 to 100 $\mu$eV, which for electron and positron sources are ultra low energies. That these ultra low kinetic energy electron and positron sources needed for this experiment are obtainable in sufficient quantities was shown in concept by Mills~\cite{Mills} (needed fast positron beam intensity of $3\times10^7$ 1/s/cm$^2$) and by experiment as discussed by Kurz~\cite{Kurz}. The possibilities would increase when instead of a $^{22}$Na source, a reactor-based positron sources~\cite{POSH},\cite{NEPOMUC} could be used where the positron yield is at least a factor of 10 larger. Another possibility is to use positron traps which can store up to $3 \times 10^{10}$ positrons per cell~\cite{Danielson} and release them in pulses.
\subsection{Focusing, shielding and flight path}
Focusing has to be done by means of a symmetric time-of-flight dependent electrostatic or magnetic lens, because the focus distance of such a lens is determined by the relative kinetic energy change of the particles passing the lens. The ability to tune the lens to the right field value will determine for a large portion the minimal attainable kinetic energy or maximal attainable deflection. An important design criterion is the wavelike structure that electrons and positrons exhibit. The De Broglie wavelength is inversely proportional to the velocity given by 
\begin{equation}
\lambda = \frac{h}{m_i v}  = \lambda_0 \frac{ v_0}{v} 
\end{equation}
where $h=6.626\times 10^{-34}$~Js, $m_i=9.109\times 10^{-31}$~kg is the electron (or positron) inertial mass, $\lambda_0=100$~nm for $v_0=7.27$~km/s.
Due to this wavelike structure of the particles, the circular apertures in the middle of the setup result in a Fraunhofer diffraction pattern at the detector plane. The most simple diffraction pattern from a circular aperture with diameter $D$ is the Airy pattern where the inner most intense fringe is called the Airy disk. This Airy disk has a diameter of 
\begin{equation}
d = 1.22 \lambda \frac{ L}{D} = 1.22 \frac{ \lambda_0 v_0 }{v} \frac{L}{D} 
\end{equation}
as long as $D >> \lambda$. Note that the Airy disk size is inversely proportional to the velocity of the particles, while the deflection is inversely proportional to the square of the velocity.

The diameter of the Airy disk should be less than the anticipated deflection (Rayleigh's criterion), hence
\begin{equation}
t =  L / v >  2.44 \frac{\lambda_0 v_0}{D \left|g_{e,p}\right|} \approx \frac{D_0 t_0}{D}   
\end{equation}
where $D_0=10$~cm and $t_0=1.81$~ms. {\it Hence, due to the wavelike nature of the particles, the minimal time-of-flight needed to obtain a sufficient resolution is inversely proportional to the diameter of the aperture.} Note that for $L=13$~m and $D=10$~cm, the velocity of the particle should be maximal $7.3$~km/s, hence its wavelength at least $100$~nm and its corresponding kinetic energy maximal $150~\mu$eV. In such a case the deflection would be minimal 16~$\mu$m. The deflection increases to 0.1~mm for particles with a kinetic energy of 25~$\mu$eV. If one would take the values used by Mills~\cite{Mills} $D=10$~cm and $L=100$~m, then the velocity of the particle should be maximal $55.2$~km/s, hence its wavelength at least $13$~nm and its corresponding kinetic energy maximal $8.7$~meV. In such a case the minimal deflection would still be only 16~$\mu$m. The deflection would however increase to 5.6~mm for particles with a kinetic energy of 25~$\mu$eV. 

\begin{figure}[t]
\begin{picture}(300,140)\put(30,0){
\put(0,10){\scalebox{0.3}{\includegraphics{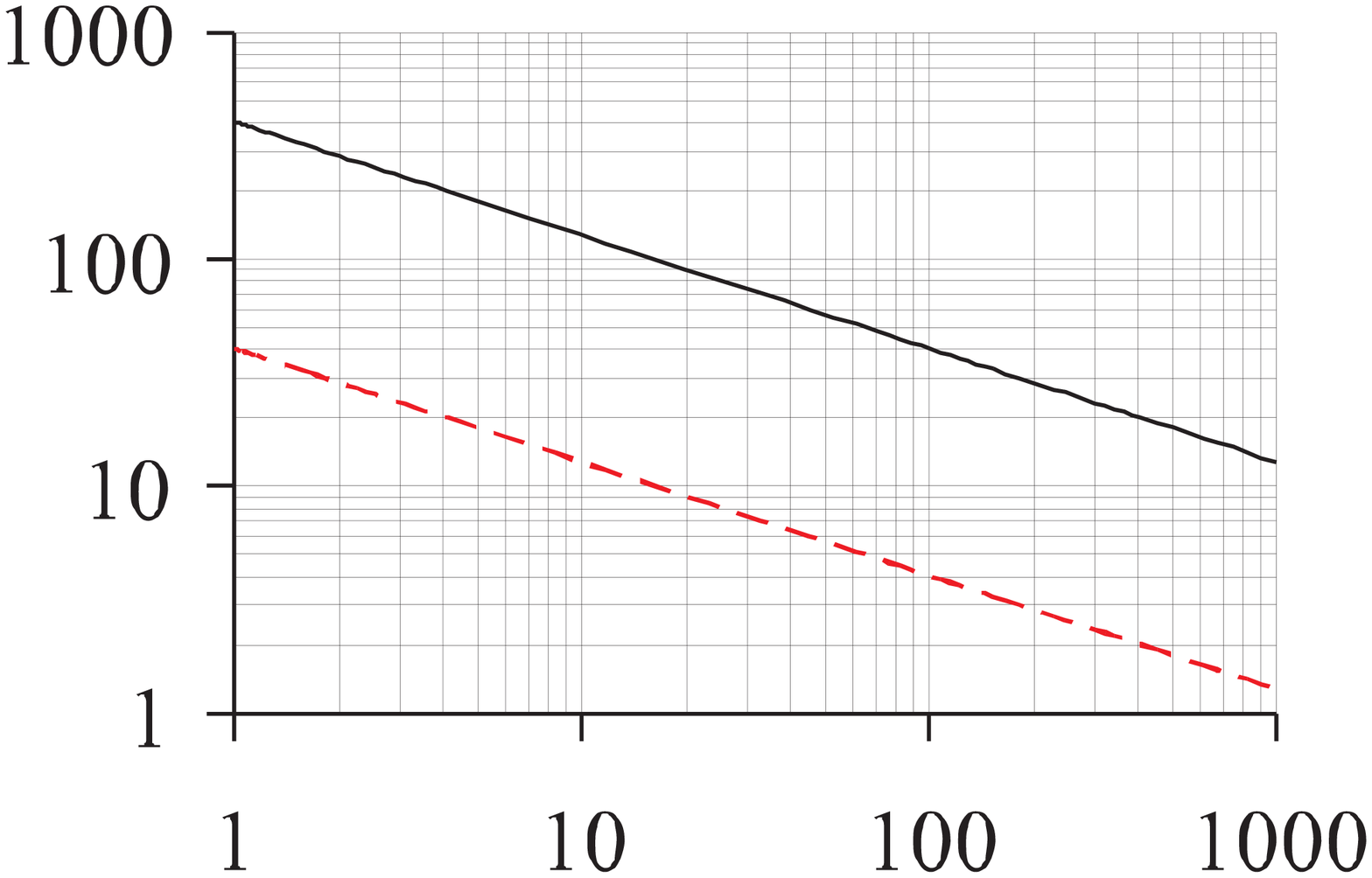}}}
\put(40,-5){Resolution diameter / $\mu$m}
\put(20,130){Minimal flight path  / m}
\put(180,10){\scalebox{0.3}{\includegraphics{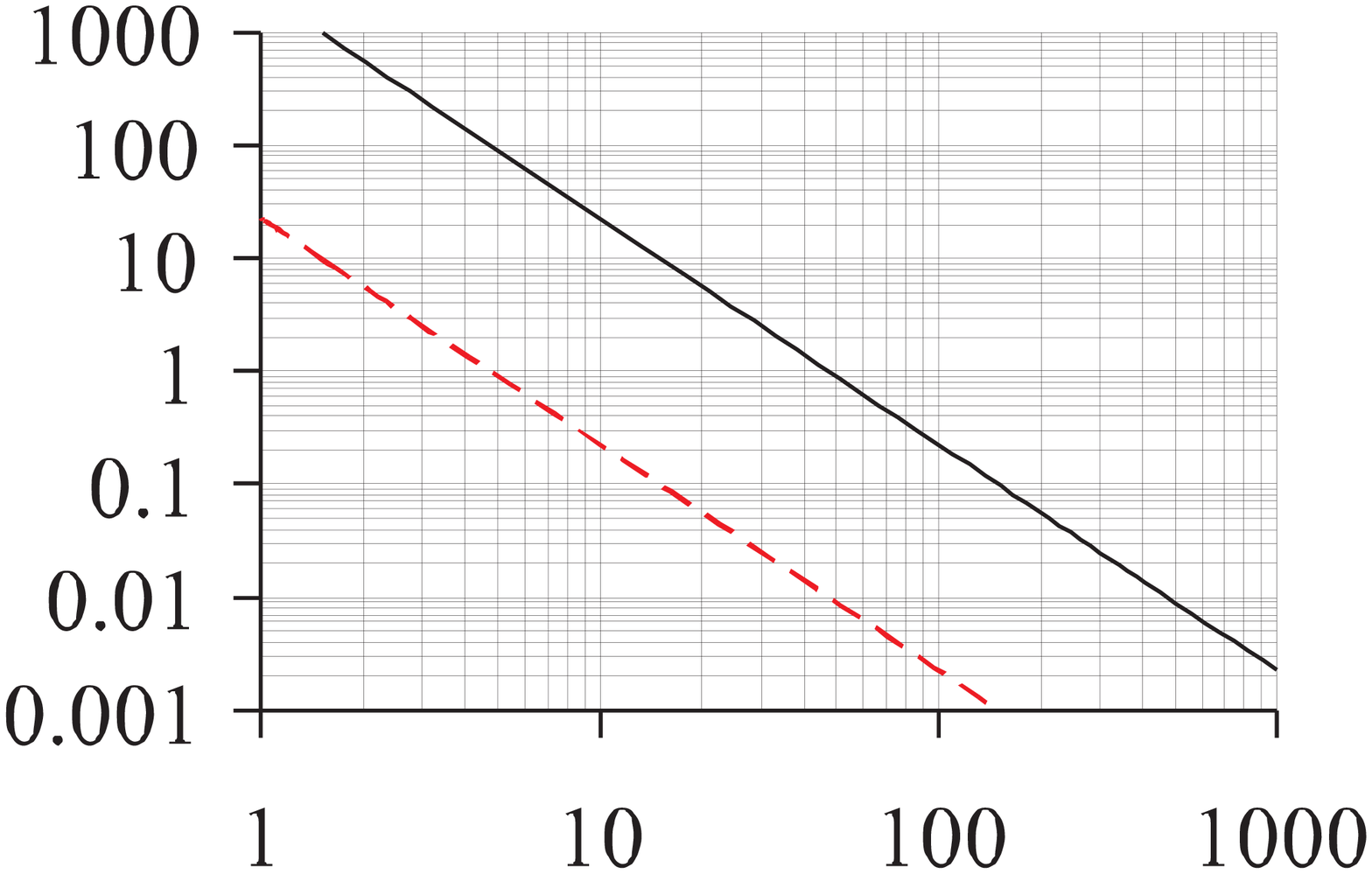}}}
\put(220,-5){Resolution diameter / $\mu$m}
\put(200,130){Kinetic Energy / meV}
}
\end{picture}
\caption{\label{fig3} Left: Graph of the maximal kinetic energy of the particles as function of aperture diameter in a gravity experiment to assure sufficient spatial resolution. Right: Graph of the minimal needed flight path as function of the same. Solid black line for $\eta = 0.001$; dashed red line for $\eta=0.01$ (see text).}
\end{figure}

In reality the source will have a finite dimension, increasing the above mentioned spot diameter. For an ideal instrument the image of the source on the detector plane and the Airy disk should have approximately the same size and be comparable to the detector resolution. In such a case the minimal needed aperture is completely determined by the needed resolution 
\begin{equation}
D_{min} = 1.73 \frac{ \lambda_0 v_0}{ \sqrt{d \left|g_{e,p}\right|}}   
\end{equation}
This also fixes the minimal needed length of the instrument as $D/L$ is between 0.1 and 0.001. The upper bound is due to limitation of the particle-optics components (par-axial approximation) and
the lower bound due to intensity limitation as the particle intensity on the detector is given by all the particles that are passed through the aperture and is proportional to $\eta^2$, hence $\eta$ cannot be made too small. If it is used that $D/L=\eta$, then the maximum velocity to obtain a sufficient resolution is given by  
\begin{equation}
v = 1.22 \frac{ \lambda_0 v_0 }{d \eta}
\end{equation}
and the corresponding maximal kinetic energy
\begin{equation}
E_{kin} = 0.74 m_i \left(\frac{ \lambda_0 v_0 }{d \eta}\right)^2     
\end{equation}
The maximal kinetic energy of the particle as function of the aperture diameter is shown in the left graph of figure~\ref{fig3} for different values of $\eta$. The corresponding minimal length of the flight path is shown in the right graph. From these graphs one can see that the choices made by Santilli and Mills to use a flight path between 10 and 100 m is a good compromise between the needed flight path (as small as possible) and the needed minimal kinetic energy (as large as possible).
A flight path as large as possible would be optimal as all other requirements relax when the flight path increases. However, the realization costs for the flight path will be roughly proportional to the square of the flight path length because for an optimal performance the diameter of the flight path has to be proportional to its length. If only the length will be made larger and not the diameter then the advantage of increasing the flight path is lost in the reduction of intensity. Hence, the optimal flight path depends on budget but probably will be between 10 and 100 m. 

Probably the most crucial part of the instrument will be the shielding of residual electric and magnetic fields. The most important components that need to be shielded sufficiently well are those resulting in a force in the same (or opposite) direction as gravity. An extensive review of all possible fields that need to be shielded is given by Darling~\cite{Darling}. His conclusion is that with the current technology it is possible to construct an adequate shielding. The way this can be done is described by Mills~\cite{Mills}. It consists of a stacked layer system of different materials cooled to a temperatures close to 4.2~K to obtain optimal shielding. 

\subsection{Surface patch potential effect}
The only remaining shielding issue is the electric potential variation in the flight path of the particle due to the inner surface of the most inner layer of the flight path tube. This inner surface consists of small crystallites exhibiting a small potential variation, these constitute the so-called patch-effect. This might cause a potential variation of about 1 $\mu$V on the axis of the flight tube. This is a reason why the inner shield must also be cooled down to liquid helium temperatures reducing the patch effects. 

A way of determining the influence of the patch effect is to estimate the optical phase differences due to potential variations over different paths from source to detector. The optical phase along a particle trajectory is given by
\begin{equation}
\psi = \frac{2\pi}{\lambda} \oint n(\vec{s}) d\vec{s}
\end{equation} 
where $n(\vec{s})$ equals the refractive index along the trajectory defined by $\vec{s}$. This refractive index is coupled to the potential by 
\begin{equation}
n(\vec{s})^2 =1 \pm \frac{2em_i\lambda^2V(\vec{s})}{h^2}
\end{equation} 
where $e=1.602\times 10^{-19}$~C is the elementary charge and $V(\vec{s})$ is the potential along the trajectory. The plus holds for electrons, the min for positrons. Variations in the potential due to the patch effect are very small, hence the variations in the refractive index can be approximated by  
\begin{equation}
\Delta n(\vec{s})= \pm \frac{e m_i \lambda^2 }{h^2} \Delta V(\vec{s})
\end{equation} 
and variations in the optical phase are directly related to variations in the potential according to
\begin{equation}
\Delta \psi = \pm 2\pi \frac{e m_i \lambda }{h^2} \oint \Delta V(\vec{s}) d\vec{s}
\end{equation} 
According to Darling~\cite{Darling} a Gaussian distributed patch effect (with root-mean-square patch potential, $\phi_{patch}$ and average crystallite size $\zeta$) on the inner surface of a long cylinder ($L>>D$) results in potential variations of $\phi_{patch}2\zeta/D$ on the axis. The line integral over these variations can be estimated by transforming the integral over a sum of $L/\zeta$ patches of length $\zeta$. The sum can be regarded as a random walk, so that the final spread in $\psi$ becomes
\begin{equation}\label{eq:phasespread}
\frac{\sigma_\psi}{2\pi} =   \frac{\lambda}{\lambda_0} \frac{\zeta}{D} \frac{\phi_{patch} \sqrt{L\zeta} }{P_0}  < 1
\end{equation} 
where
$P_0 = h^2/(2e m_i\lambda_0) = 1.5\times 10^{-11} Vm$. To be able to get a good focus this variation in optical phase should be much smaller than $2\pi$. Note that the variation is proportional to the wavelength, which clearly favors faster particles.

For the optimal resolution setup of the previous section this condition puts a limit on the ratio between $D$ and $L$
 \begin{equation}
\eta < 13.5 \left( \frac{ d_0 }{d }\right)^{5/2} \left(\frac{\zeta_0}{\zeta}\right) \left(\frac{\phi_{0}\zeta_0}{\phi_{patch}\zeta }\right)^2
\end{equation} 
where $d_0=100~\mu$m,  $\zeta_0 = 1~\mu$m, $\phi_{0}= 1~\mu$V. This is completely determined by the required resolution and the patch potential distribution. If a resolution of 100~$\mu$m is required and $\eta$ would be between 0.001 and 0.1, then for $\zeta = 1~\mu$m, $\phi_{patch}$ has to be less than 100 to 10~$\mu$V, which is perfectly feasible~\cite{Labaziewicz}.

According to equation~(\ref{eq:phasespread}) the spread in optical phases close to the cylinder axis is proportional to the wavelength. This explains why the vertical flight path as used by Witteborn~\cite{Witteborn2} is much more sensitive to the patch effect than the horizontal flight path considered here. Take $\lambda=\lambda_0v_0/\sqrt{gL}$ (the average wavelength for a particle just reaching the top of the flight path), then for $L=1$~m, $D=4$~cm and $\zeta =1~\mu$m, $\phi_{patch}$ has to be less than 250~nV at least a factor of 400 smaller. Darling~\cite{Darling} takes $\zeta << 1$~nm and $\phi_{patch}=0.01$~V, as limit which corresponds to a variation of the optical path phase of $\sigma_\psi << 2.4 \pi$. Hence, both approaches give similar results. 

Equation~(\ref{eq:phasespread}) can be rewritten as function of the total deflection of the particle beam
\begin{equation}\label{eq:phasespread2}
\frac{\sigma_\psi}{2\pi} =   \frac{ 2 e }{h } \sqrt{\frac{2 \Delta z} { \left|g_{e,p}\right| }} \frac{\phi_{patch} \zeta}{D}  \sqrt{\frac{\zeta}{L}} 
\end{equation} 
This is independent of the particle properties. Hence, {\it for a required given deflection in the proposed experiment, the influence of the patch potential effects does not depend on the type of particle used}.

In view of the relatively large kinetic energies involved in this horizontal flight path experiment with regard to the Witteborn experiment~\cite{Witteborn} and the implicit determination of the average kinetic energy by means of the time-of-flight method, the influence of the patch-effects will be much reduced. 

This also relaxes the requirements on the vacuum pressure quality to about $10^{-8}$~Pa as the time-of-flight is at least a factor of 100 shorter and the main effect it has on the results is a reduced intensity at the detector. 

\subsection{Electron and positron detection}
The preferable detector should be a linear position sensitive detector that can detect both electrons and positrons. The spatial resolutions should be in the order of 100~$\mu$m and the time resolution of the order of 0.1~ms with an efficiency as high as possible. These are moderate requirements and can be met by for instance micro channel plates~\cite{MCP1},\cite{MCP2},\cite{Alpha} or linear CMOS detectors~\cite{CMOS}. 

\section{Conclusions}
The above shows that with current technology it is perfectly feasible to perform the long awaited experiment to compare the gravitation on electrons and positrons as suggested by R.M. Santilli~\cite{Santilli02} almost two decades ago. The largest challenge will be the adequate shielding of the flight path to acceptable levels by means of a supercooled vacuum tube.
\section*{Acknowledgments}
This research project was partially supported by a Research Grant from the R. M. Santilli Foundation Number RMS-AM-4673rs82810.

\end{document}